\documentstyle[psfig]{mn}
\def\etal{{\it et al. }}
\def\kms{\rm ~km~s^{-1}}
\def\gsim{ \lower .75ex \hbox{$\sim$} \llap{\raise .27ex \hbox{$>$}} }
\def\lsim{ \lower .75ex \hbox{$\sim$} \llap{\raise .27ex \hbox{$<$}} }

\input{epsf}

\title{Ram pressure stripping of spiral galaxies in clusters} 
\author[Abadi et al. ]
       {Mario G. Abadi$^{1,2}$, Ben Moore$^1$ \& Richard G. Bower$^1$\\
        $^1$Department of Physics, University of Durham, South Road, DH1
        3LE, Durham, UK\\
	$^2$Observatorio Astron\'omico, Universidad Nacional de C\'ordoba,
	Laprida 854, 5000 C\'ordoba, Argentina}

\date{\today}

\begin{document}
\maketitle

\begin{abstract}

We use 3-dimensional SPH/N-BODY simulations to study ram pressure
stripping of gas from spiral galaxies orbiting in clusters.  
We find that the analytic expectation of Gunn \& Gott (1972) relating
the gravitational restoring force provided by the disk to the ram
pressure force, provides a good approximation to the radius that gas
will be stripped from a galaxy.  However, at small radii it is also
important to consider the potential provided by the bulge component.
A spiral galaxy passing through the core of a rich cluster such as
Coma, will have its gaseous disk truncated to $\sim 4$ kpc, thus
losing $\sim 80\%$ of its diffuse gas mass. The timescale for this to
occur is a fraction of a crossing time $\sim 10^7$ years. Galaxies
orbiting within poorer clusters, or inclined to the direction of motion
through the intra-cluster medium will lose significantly less gas.
We conclude that ram-pressure alone is insufficient to account for the
rapid and widespread truncation of star-formation observed in cluster 
galaxies, or the morphological transformation of Sab's to S0's that is
necessary to explain the Butcher-Oemler effect.

\end{abstract}

\begin{keywords}
galaxies: ram-pressure.
\end{keywords}

\section{Introduction}


There is a long standing debate concerning the effect of environment
on galaxy morphology. The pioneering work of Butcher \&
Oemler (1978, 1984) first demonstrated that distant clusters contained a 
far higher fraction of blue galaxies than their local counter-parts.
Subsequent work has established that many of the red galaxies in these
clusters have spectral signatures of recent star formation 
(Dressler \& Gunn, 1983, Couch \& Sharples 1987, van Dokkum et al.,
1998, Poggianti et al., 1998). 
The most recent advances have been made with the 
Hubble Space Telescope that allows the morphology of the distant
galaxies to be directly compared with the properties of their nearby
counter-parts. The studies of Dressler \etal (1997) and Couch \etal (1998)
suggest that the predominant evolutionary effects are that the distant clusters
have a substantial deficit of S0 systems compared to nearby systems, and
at lower luminosities they contain primarily Sc-Sd spirals
compared with the large population of dwarf spheroidals in present
day clusters. 

Many authors have suggested that the predominance of
early-type S0 galaxies in local clusters is due to a mechanism that
suppresses star formation in these environments leading to a
transformation of galaxy morphology.  Comparison of the galaxy
population of local and distant clusters provides the strongest
evidence for this.  This leads to the natural conclusion that the
primary effect of the cluster environment is to transform luminous 
spiral galaxies into S0 types through suppression of their star formation.

A key ingredient in the explanation of the Butcher-Oemler effect is 
the rate at which `fresh' galaxies are supplied from the field into
the cluster environment.
The differences in the fractions of blue, or actively star forming, galaxies
between the local and distant clusters may result either from an increase
the the general level of star formation activity at higher redshift
(eg., Lilly \etal 1996, Cowie \etal 1997), 
or might result from a different level of infall
between local and distant clusters (Bower, 1991, Kauffmann 1996).

Several mechanisms have been proposed that may be capable of
explaining the transformation of galaxy morphology in dense
environments. Ram-pressure stripping has been a long standing
possibility, dating from the analytic work of Gunn \& Gott (1972), and
a mechanism that has been cited in over 200 published abstracts. As a
galaxy orbits through the cluster, it experiences a wind due to its
motion relative to the diffuse gaseous intra-cluster medium
(ICM). Although the ICM is tenuous, the rapid motion of the galaxy
causes a large pressure front to build-up in front of the
galaxy. Depending on the binding energy of the galaxy's own
interstellar medium, the ICM will either be forced to flow around the
galaxy, or will blow through the galaxy removing some or all of the
diffuse interstellar medium.  
Related mechanisms to ram-pressure are thermal evaporation of the
interstellar medium (Cowie \& Songaila 1977) and viscous stripping of
galaxy disks (Nulsen 1982).  These occur even when the ram-pressure is
insufficient to strip the gas disk directly: turbulence in the gas
flowing around the galaxy entrains interstellar medium resulting in
its depletion.

If ram-pressure or viscous stripping is effective at removing
gas, then cluster spirals should have truncated disks deficient in HI.
Observational evidence for this is marginal. Some galaxies show clear
evidence for stripping, e.g. NGC 4522 (Kenny \& Koopmann 1998), 
UGC 6697 (Nulsen 1982) or several Virgo cluster galaxies (Cayatte 1994).
However, a larger survey of 67 cluster galaxies 
showed no evidence of these effects (Mould \etal 1995).

Interactions between galaxies are another possible agent for promoting
morphological transformation. However, strong interactions that
lead to galaxy merging, are unlikely to be an effective
mechanism in virialised clusters of galaxies, since the relative velocity of
galaxies is too high for such encounters to be frequent ({\it i.e.}
Ghigna \etal 1998). Moore \etal (1996) examined the effects of rapid
gravitational encounters between galaxies or with the lumpy potential
structure of clusters.  This mechanism has been termed galaxy
`harassment' and is highly effective at transforming fainter Sc-Sd
galaxies to dSph's and even tidally shredding LSB galaxies.  
Although this mechanism can account for the observed evolution of lower
luminosity galaxies in clusters, the concentrated potentials of
luminous Sa-Sb galaxies help to maintain their stability (Moore \etal
1999), although their disks are substantially thickened.

A final mechanism that should not be over-looked is the truncation
of star formation through the removal of the hot gas reservoir that
is thought to surround galaxies (Larson \etal 1980, Benson \etal 1999). 
In clusters any hot diffuse material
originally trapped in the the potential of the galaxies halo becomes
part of the overall ICM. The galaxy (with the possible exception of the
central dominant galaxy) cannot supplement its existing ICM and thus 
is doomed to slowly exhaust the material available for star formation. 
This mechanism is the only environmental mechanism currently embedded
into hierarchical galaxy formation codes 
(eg., Kauffmann \& Charlot 1998, Baugh \etal 1998);
however, it appears unable to adequately reproduce the star formation
histories of real cluster galaxies, since the spectroscopic studies
require that star formation is suppressed on far shorter timescales
(eg., Barger \etal 1996, Poggianti \etal 1998).

In this paper, we revisit ram-pressure stripping as a mechanism for the
removal of gas from cluster galaxies and thus rapidly suppressing the
star-formation rate. In particular, we use fully
3 dimensional SPH simulations to compare with the analytic estimate
of Gunn \& Gott, and to investigate the effect of differing galaxy
infall velocities, inclinations and cluster gas densities.
Our main motivation is to investigate whether ram
pressure could be effective in clusters less rich than the Coma 
cluster, and to determine whether the stripping effect is limited
only to the outer-part of the disk or whether the effects can propagate
inwards. It is also of interest to determine the timescale on which
the stripping should occur, since rapid truncation of star formation
appears to be one of the key signatures of the spectroscopic
Butcher-Oemler effect. 

Relatively little numerical work has been performed to examine the effects of ram
pressure, especially 3 dimensional simulations {\it c.f.} Farouki \& Shapiro (1980).
Balsara \etal (1994) performed several high resolution simulations of the gas
stripping process using an Eulerian code, but again using a restricted 2-dimensional
version. As well as being able to study the rich structure of the gas ablation process,
these authors found that the galaxy could accrete gas from 
the downstream side of the flow. 
We note that Kundic \etal (1993) also reported a 
preliminary investigation of this problem using a smoothed particle 
hydrodynamic (SPH) code.

The structure of this paper is as follows. In Section 2 we discuss the 
parameters of the galaxy model that we shall use and make predictions for
the radius that gas will be stripped via ram pressure. Techniques and results
of numerical SPH simulations are presented in Section 3, which are discussed
in Section 4, along with the shortcomings of ram pressure stripping as the
mechanism behind explaining the Butcher-Oemler effect.

\section{The galaxy model}

We construct an equilibrium galaxy model designed to represent the
Milky Way, using the techniques described by Hernquist (1993). The
model has a stellar and a gaseous disk, halo and bulge components.
The bulge is spherical and has a mass density profile of the form:

\begin{equation}
\rho_b(r)=\frac{M_b}{ 2 \pi r_b^2} \frac{1} { r ( 1+r/r_b)^3 },
\end{equation}

where $r_b$ is the scale length and $M_b$ is the mass.
The model has a dark matter halo with 
density given by the following truncated profile:

\begin{equation}
\rho_h(r)= \frac{M_h}{2 \pi^{3/2}} \frac{\alpha}{r_t r_h^2} \frac{exp(-r^2/r_t^2)}{(1+r^2/r_h^2)}
\end{equation}

where $r_h$ is the core radius, $r_t$ is the truncation radius and $M_h$ is the mass.
The mass normalisation requires that the constant:
\begin{equation}
\alpha=1/\{1-\pi^{1/2} q \; exp(q^2)[1-erf(q)]\}
\end{equation} 

where $erf(q)$ is the error function and $q=r_h/r_t$.
The disk is axisymmetric and is composed of both stars and gas. 
Its mass density profile is an exponential of the form:

\begin{equation}
\rho_d(R,z)={M_d \over{4 \pi R_d^2 z_d}}  exp(-R/R_d) sech^2(z/z_d)
\end{equation}

where $R_d$, $z_d$ and $M_d$ are the cylindrical scale length, the
vertical thickness and mass, respectively.  We will replace the
subscript $d$ by $s$ to refer to the stellar disk, or $g$ to refer to
the gaseous disk.  

\begin{figure}
\epsfxsize=\hsize\epsffile{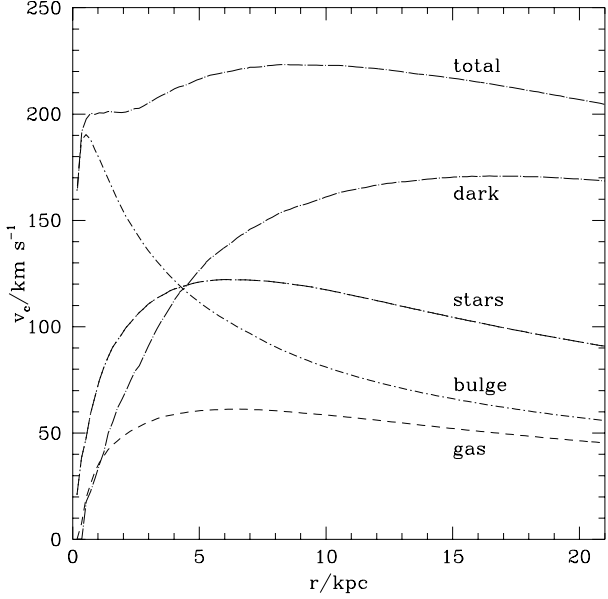}
\caption{
The contribution from each component of the model galaxy to the
rotational velocity of the disk. 
}
\label{fig:vc}
\end{figure}

The characteristic length scales of each component are listed in Table
1 and we plot the contribution to the rotational
velocity of the disk provided by each component in 
Figure~\ref{fig:vc}.  Dark matter begins
to dominate the baryonic components beyond $\sim 10$ kpc, and the
maximum rotational velocity of the disk is $220 \kms$.  Adopting a B
band mass to light ratio of 2, the central surface brightness of the
model galaxy is $\sim 21$ mags arcsec$^{-2}$.


\section{Analytic solution}

We have applied the ideas of Gunn \& Gott (1972) to the galaxy
model in order to obtain analytic estimates of the radius beyond which the
gas will be stripped, 
$R_{str}$, when the galaxy is moving through 
the intracluster medium (ICM).
These authors stated that the gaseous disk will be removed if the ram
pressure of the ICM is greater than the restoring gravitational force
per unit area provided by the galaxy's disk ({\it c.f.} Sarazin
1986). In this case, the ram pressure is $P = \rho_{icm} v^2$, where
$v$ is the velocity of the galaxy with respect to the ICM and
$\rho_{icm}$ is the gas density of the ICM.  The restoring
gravitational acceleration of a particle orbiting in the galaxy is
${\partial \phi}/{\partial z}$, 
where $z$ is the coordinate
perpendicular to $v$.  The total gravitational potential, $\phi$, of the
galaxy can be obtained solving the Poisson equation $\nabla^2
\phi(R,z) = 4 \pi G \rho(R,z) $ for each component separately 
and summing.
In the case of a face-on passage, we have for the bulge,
\begin {equation}
\frac {\partial \phi_b}{\partial z}(R,z)= \frac {G M_b} {(r+r_b)^2} \frac {z} {r}
\end{equation}
and for the halo,
\begin {equation}
\frac {\partial \phi_h}{\partial z}(R,z)= \frac {2 \alpha G M_h} {\pi^{1/2} r^2}  \frac{z}{r} 
\int_0^{r/r_h} \frac {x^2 exp(-x^2)}{x^2+q^2} dx.
\end{equation}

The analytical solution of the Poisson equation for the disk is not so
straightforward as for the spherical components. Binney \& Tremaine (1987) use
separation of variables to solve this problem for the case of an
infinitely thin disk with a surface density

\begin{equation}
\sigma_d(R)=\int_{-\infty}^{\infty}\rho_d(R,z) dz= \frac {M_d}{2 \pi R_d^2} exp(-R/R_d) .
\end{equation}

We have adopted this approximation and used their formula (2-167) in order to compute
the restoring gravitational acceleration for the disk:

\begin {equation}
\frac {\partial \phi_d}{\partial z}(R,z)= G M_d \int_0^{\infty} \frac{J_0(kR)exp(-k|z|)}{[1+(kR_d)^2]^{3/2}} k dk.
\label{equation:thin}
\end{equation}
where $J_0(x)=\pi^{-1}\int_0^\pi cos(x\;sin\theta) d\theta$ is the
Bessel function of the first kind of order zero.  The stripping
radius $R_{str}$, can then be computed by solving the equation

\begin{equation}
\frac {\partial \phi}  {\partial z} (R_,z) \sigma_g(R) = \rho_{icm} v^2, 
\label{equation:ram}
\end{equation}

where the left hand side is the total restoring gravitational force
per unit mass of the model and the right hand side is the ram
pressure. At a given radius, $R$, the restoring gravitational force per
unit mass is a function of the coordinate $z$. This force is 
maximum at $z=0$ therefore in 
order to completely remove the gas from the galaxy the ram pressure
must be greater than this value.

\begin{figure}
\epsfxsize=\hsize\epsffile{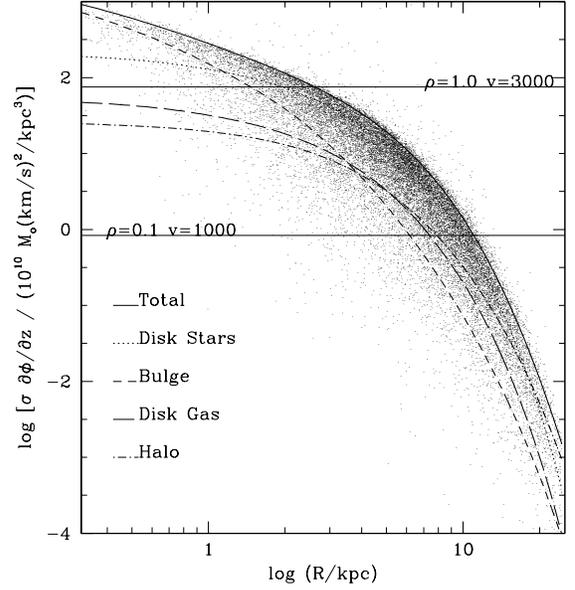}
\caption{ The component of the force in the direction $z$ 
perpendicular to the velocity
flow, as a function of cylindrical radius for each component.
The curves show the total contribution for each component of the model
galaxy whilst the dots show values calculated at the positions of each individual
particle. The horizontal lines show the ram pressure force for
representitave velocities and ICM densities.
}
\label{fig:mnumford}
\end{figure}

In Figure~\ref{fig:mnumford} we plot both sides of 
Equation~\ref{equation:ram} (solid lines),  
for the parameters quoted in Table~\ref{table:model}, 
which were chosen to represent observational values 
for an Sb type spiral galaxy like the Milky Way. We also
show the contribution of each component to the maximum restoring force
per unit mass as a function of the radius.
The dots show the value of the total
gravitational force per unit area computed directly from all particles
of the N-body realisation.  The agreement between the analytical
estimates (solid curved line) and the envelope of values computed from the particles
(dots), demonstrate the validity of using Equation~\ref{equation:thin}.
\begin{table}
\begin{center}
\begin{tabular}{l|r|r|l|l|l}
\hline
            &  $N$   &  $M_{total}$  & $m_p$ & $[L]$ \ kpc & $\epsilon$ \ kpc \\
\hline
ICM         & 16000 &  0.24 & 1.47e-5    &               &  0.75   \\
Disk(gas)   &  8000 &  1.12 & 1.40e-4    & 3.5 0.35      &  0.28   \\
Disk(stars) &  8000 &  4.48 & 5.60e-4    & 3.5 0.35      &  0.28   \\
Bulge       &  2500 &  1.68 & 6.72e-4    & 0.5           &  0.21   \\
Halo        &  8000 & 21.28 & 2.66e-3    & 3.5 24.5      &  1.40   \\
Total       & 42500 & 28.58 &            &               &         \\

\hline
\end{tabular}
\caption{
The main parameters of each component of the model spiral galaxy. 
$N$ is the number of particles.
$M_{total}$ is total mass of each component (for the ICM, this is quoted assuming
the value for a density equal to the core of the Coma cluster) and 
$m_p$ is the mass per particle for each component, all expressed in units of
$10^{10}M_\odot$. 
$[L]$ is the characteristic length scales of the different components 
(for the disk we quote $R_d$ and $z_d$; for the 
halo we quote $r_h$ and $r_t$).
$\epsilon$ is the gravitational softening.
}
\label{table:model}
\end{center}
\end{table}

The ram-pressure values for 2 different ICM densities and relative
velocities are shown as horizontal lines in 
Figure~\ref{fig:mnumford}.  
These may be representative of galaxies passing through
the core of the Coma and Virgo clusters.
The predicted radius of the final stripped gas disk,
$R_{stp}$, is given by the intersection of the horizontal line with
the total restoring force per unit area (solid curved line), roughly 3 kpc and
10 kpc for the values illustrated here.  For $R<R_{stp}$ the restoring
gravitational force per unit mass is greater the the ram-pressure and
the gas remains bound to the galaxy. On the contrary, for $R>R_{stp}$
the ram-pressure overcomes the gravitational force and the gas can be
stripped.

Figure~\ref{fig:mnumford} demonstrates that the main
contribution to the total restoring gravitational force comes from the
stellar disk, although in the central parts of the galaxy the bulge
becomes important. For this model, the bulge contributes 30\% of the
disk mass and dominates the vertical potential in the central 2 kpc.
The halo provides a negligible contribution within 10 kpc, but begins
to dominate on scales $\gsim 20$ kpc.

One interesting and straightforward application of this model is the
comparison of $R_{stp}$ with the $H_I$ observational data available for
galaxies in clusters. Cayatte \etal (1994) analyse the surface
brightness of 17 bright spirals in the Virgo cluster. They
divided the sample into 4 subsamples according to the shape of the
surface brightness profile and conclude that ram-pressure is the main
reason of gas removal in subsample III (3 galaxies).  They also
present a list of isophotal $H_I$ and optical diameters for these
galaxies.  

\begin{figure}
\epsfxsize=\hsize\epsffile{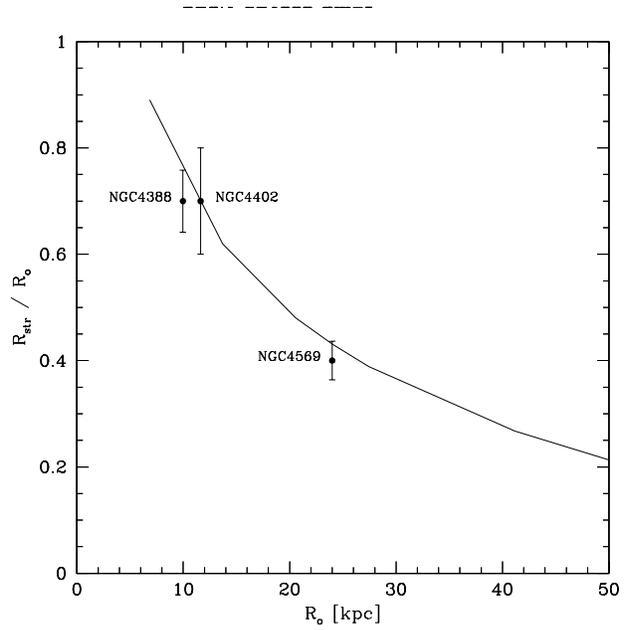}
\caption{Dependence of the stripping radius $R_{stp}$, as a function of the 
the disk scale length $R_d$.
}
\label{fig:lsb1}
\end{figure}

We solved Equation~\ref{equation:ram} for a ram-pressure
corresponding to $\rho_{icm}=0.1 \rho_C$ and $v=1000 
\kms$, with
different values of the galaxy's scale length $R_d$. In order to
compare directly with the observations we define an ``optical'' radius
$R_o$ as the size of the stellar disk.
Then, we scale linearly from
$R_d$ to $R_o$.  For $R_d=3.5$ kpc this value is $R_o=24$ kpc and at
this position the density has decreased by a factor $\sim 10^{-3}$.  In
Figure~\ref{fig:lsb1} we show the ratio of the stripping radius to
``optical'' radius $R_{stp}/R_o$ as a function of $R_o$ for the model
(solid line).  The filled circles show the observed radii
obtained by Cayatte el al. (1994) for their galaxies and we find
reasonable agreement between the model and the data.

\section{Numerical simulations }

Hydro-dynamical simulations of the ram-pressure require enough spatial
resolution to follow the interaction between the ``cold'' disk gas and
the hot ICM. If the ICM particles are too massive, then they will
punch holes in the gas disk like bullets and the flux of particles
against the disk will be dominated by shot noise.  With current
computational resources, simulations that attempt to capture the full
cosmological context of the formation of disks and their subsequent evolution
within a cluster environment will be 
completely dominated by numerical effects. 

Ideally, gas particles will flow onto the disk, imparting a
significant fraction of their momentum as their motion is halted by
the disk gas. Too hot to accrete onto the disk, they will be
forced to flow around and re-join the ICM.  To achieve this
spatial resolution, the ICM particles should have a mass that is at
least as small as that of the disk gas. The limitation in the number of gas
particles that SPH-codes can handle makes it impossible to follow the
evolution of a galaxy through an entire cluster. 

For example, the mass of gas inside the Abell 
radius $r_A=1.5 h^{-1} {\rm Mpc}$ 
for the Coma cluster is $M_{icm}=5 \times 10^{13} h^{-5/2}
M_{\sun}$ (White et al. 1993).  On the other hand, the $H_I$ component of a
massive galaxy is $M_g \sim 10 ^{10} M_{\sun}$ (Canizares et al. 1986,
Young et al. 1989) so that the ratio between the number of gas particles in
the ICM and the galaxy would be $\sim ~10^{4}$.  With just $10^5$ SPH
particles, a galaxy passing pericenter will encounter of order 10\--100
gas particles.  These will detonate the disk like nuclear explosions
leaving large holes and creating a large artificial drag. To suppress
this effect, a minimum ICM gas mass equal to the disk particle mass is
necessary, requiring $N \sim 10^{7-8}$ gas particles for the ICM.

To avoid this problem we simulate only the passage of the galaxy
through the cluster core, where $\rho_{icm}$ and $v$ are maximum and
the ram-pressure stripping is most effective.  We represent the ICM as
a flow of particles along a cylinder of radius $R_{cyl}=30$ kpc and
thickness $z_{cyl}=10$ kpc.  The axis of the cylinder is oriented in
the $z$-direction, perpendicular to the plane of the galaxy in the
face-on case.  We also carry out simulations in which the galaxy is
passing edge-on and inclined at $45^o$ to 
the direction of motion through the ICM, for which we use a box of size
$60 {\rm kpc} \times 60 {\rm kpc} \times 10$ kpc.  Initially, 
we randomly distributed 
$N_{ICM}=16000$ gas particles inside the cylinder ($N_{ICM}=20000$ for
the box) with a density $\rho_{icm}$ and a temperature $T$.  We have
chosen the temperature $T=8$ keV and the density to range from the
central density of a cluster like Coma $\rho_{icm}=\rho_C \equiv
5.64\times 10^{-27} h_{50}^{1/3} g \; cm^{-3}$ (Briel, Henry \&
Bohringer 1992) to the density of a cluster like Virgo $\rho_{icm}=0.1
\rho_C$.  (Throughout this paper we have adopted a value of the Hubble
constant of $H_o=50 \; {\rm km s}^{-1} \; {\rm Mpc}^{-1}$).

In order to represent the passage of the galaxy through the ICM we
give all gas particles in the cylinder an initial velocity $v$. We
also carried out a test simulation in which we increased the density
of ICM particles from an initial value of zero, such that the galaxy
feels a gradual increase in pressure, rather than a sudden shock.  The
final stripping radius was the same as in the case of an instantaneous
wave of particles of the full density. This allows us to save an
important fraction of computational time.

We have used the TREE-SPH code developed and
kindly made available by Navarro \& White (1993).  We have modified
this code in order to include periodic boundary conditions for ICM gas
particles that leave the cylinder or the box.  Each particle
that leaves the cylinder or the box at $z_{cyl}/2$, is re-entered at
$-z_{cyl}/2$.  We also apply reflecting boundaries conditions for
particles that leave the cylinder edges at $x^2+y^2=R_{cyl}^2$.  In
Table~\ref{tab:simulation} we list the main characteristics of the
simulations.  
The code has individual timesteps that are typically $\sim 10^4$ years and
we run each simulation for more than $10^8$ years.

\begin{figure}
\epsfxsize=\hsize\epsffile{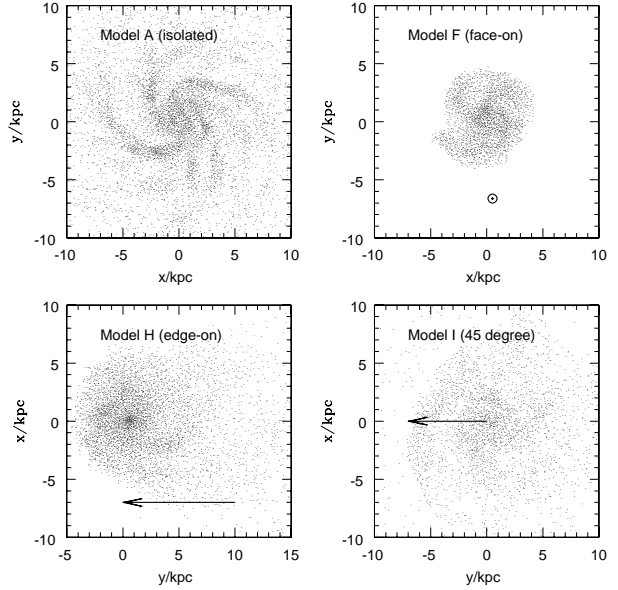}
\caption{
The positions of the gas particles plotted in the $x-y$ plane (face-on) 
after $\sim 10^8$ years of evolution. 
The arrows indicate the direction of motion (run A is the model in isolation).
}
\label{fig:xy}
\end{figure}

\begin{figure}
\epsfxsize=\hsize\epsffile{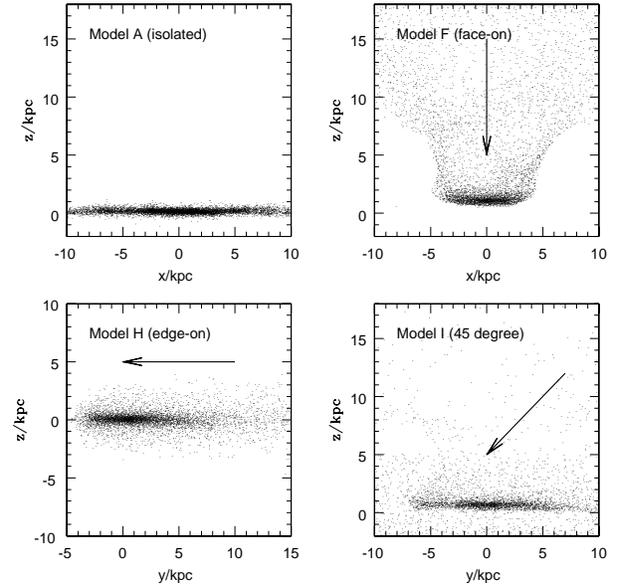}
\caption{The positions of the gas particles plotted in the $x-z$ plane 
(edge on) after $\sim 10^8$ years of evolution.
The arrows indicate the direction of motion (run A is the model in isolation).
}
\label{fig:xz}
\end{figure}

In Figure~\ref{fig:xy} and ~\ref{fig:xz} we show the
projected distribution of disk gas particles at the final output
($x-y$ plane and $x-z$ plane, respectively) for four of the simulations.  
Run A is the model galaxy in isolation, runs F, H and I are face-on, edge-on
and inclined $45^o$ to the direction of motion.
At the final time, the distribution of stars and dark matter particles
remain very similar to the initial conditions, whereas the gas
distribution is strongly modified by the ram-pressure.

\begin{figure}
\epsfxsize=\hsize\epsffile{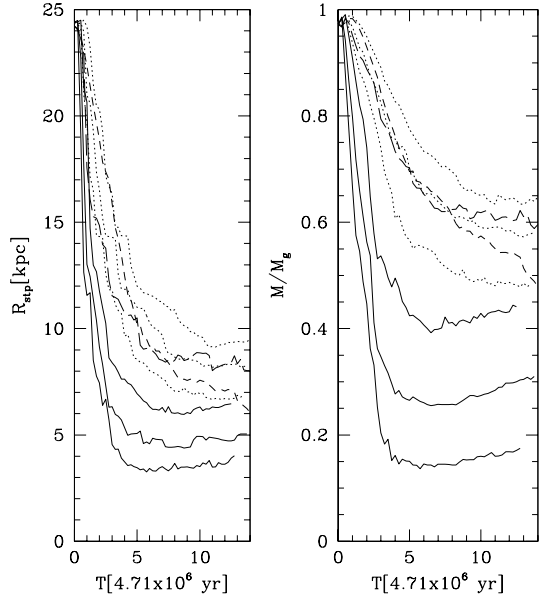}
\caption{
Evolution of the radius and mass of the gas disk as a function 
of time. Both rapidly converge on a timescale $\sim 5 \times 10^7$ years.
>From top to bottom, the dotted and dolid lines show the effects
of stripping for velocities of 1000, 2000 and $3000 \kms$ for 
$\rho=0.1\rho_{coma}$ and $\rho=1.0\rho_{coma}$.
The short and long dashed lines are for the edge-on
and $45^o$ simulations with a velocity of $2000 \kms$ and 
$\rho=1.0\rho_{coma}$.
}
\label{fig:rcutmass}
\end{figure}

In Figure
~\ref{fig:rcutmass} we show the evolution of the radius
$R_{stp}$, and the fraction of gas mass that remains inside this
radius. We estimate $R_{stp}$ as the radius of the most distant
gaseous disk particle from the center, and the mass is calculated using the
disk particles inside a cylinder of
radius $R_{stp}$ and thickness of 1 kpc. The dotted and solid curves
correspond to simulations B to G (face-on) from top to bottom,
respectively, i.e., monotonically increasing the amount of
ram-pressure. The short dashed line corresponds to simulation H (edge-on)
and the long dashed line to simulation I (inclined $45^o$).

\begin{figure}
\epsfxsize=\hsize\epsffile{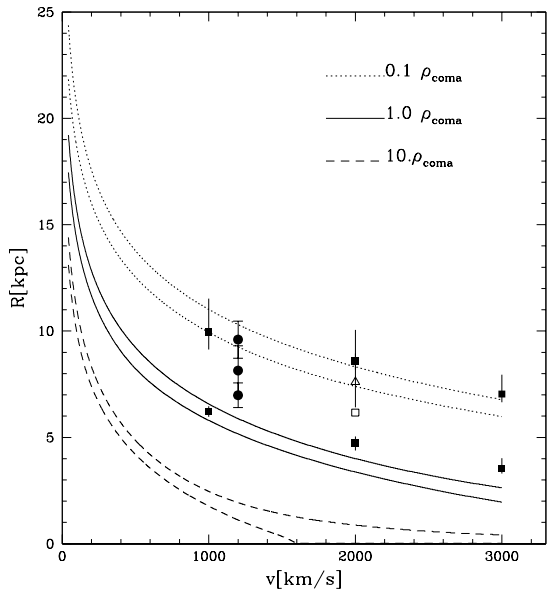}
\caption{The size of the stripped gas disk, $R_{stp}$ 
as a function of the velocity flow for 
different cluster gas densities. The filled squares are from our
simulations of face on encounters, the open square and triangle are
for edge-on and $45^o$ encounters. 
}
\label{fig:forcebs}
\end{figure}

In Figure~\ref{fig:forcebs} we have plotted the stripping radius,
$R_{stp}$, as function of the velocity $v$ of the ICM. The curves show
the analytical solution of equation~\ref{equation:ram} for different
values of density: $\rho_{ICM}=0.1 \rho_C$ (dotted line),
$\rho_{ICM}=\rho_C$ (solid line) and $\rho_{ICM}=10.0 \rho_C$ (dashed
line). For each ICM density, the upper curve shows the solution taken
into account the restoring gravitational force of all components of
the galaxy, and the lower corresponds to including just the stellar
disk. The solid squares are the values measured from the numerical
simulations of the face on passages, whilst the open square denotes
the edge-on simulation.  The open triangle shows the $45^o$ simulation
which is intermediate between face and edge-on.  We also show as solid
circles the observational radius from Cayatte et al. (1993) for 3
galaxies in the Virgo cluster, assigned a velocity of $1200 \kms$.

\begin{table}
\begin{center}
\begin{tabular}{l|l|l|l}
\hline
  run & 	$\rho/\rho_C$ &  $v$[km/s] &  flow \\
\hline
        A &     0.0 &      0 &  isolation  \\
	B & 	0.1 &	1000 &	face-on  \\	
	C & 	0.1 &	2000 &	face-on  \\	
	D & 	0.1 &	3000 &	face-on  \\	
	E & 	1.0 &	1000 &	face-on  \\	
	F & 	1.0 &	2000 &	face-on  \\	
	G & 	1.0 &	3000 &	face-on  \\	
	H & 	1.0 &	2000 &	edge-on  \\	
	I & 	1.0 &	2000 &	45$^o$  \\	
\hline
\end{tabular}
\caption{
Main characteristics of the simulations. Column 1 is the name of the run,
$\rho/\rho_C$ is the density of the intra-cluster medium in units
of the central density of Coma cluster,
$v$ is the velocity of the flow of the intra-cluster medium in $\kms$,
flow is the orientation of the galaxy with respect to the velocity of
the intra-cluster medium.
}
\label{tab:simulation}
\end{center}
\end{table}

\section{Discussion}

The motivation for this paper was to examine the effectiveness of
ram pressure stripping at removing the reservoir of cold gas from spiral
galaxies. In particular, could ram pressure be the key mechanism behind
the Butcher-Oemler effect by rapidly truncating star-formation in
cluster galaxies?

In the introduction, we outlined how a simple explanation of the 
Butcher-Oemler effect might work. New galaxies are supplied to the
dense cluster environment from the field. The evolution of the rate
of this supply is well described by numerical models for the evolution
of gravitational structure, or their analytical approximations 
({\it eg} Kauffmann 1996). Another important factor is the level of 
star formation activity in the galaxies before they feel the influence
of the cluster. It is generally believed that star formation levels
are higher in the intermediate redshift universe than locally 
(Lilly \etal 1996, Cowie \etal 1997, Steidel \etal 1998).

The second ingredient of the explanation is the effect of the 
cluster environment on the  evolution of the galaxies' star 
formation rates. A general decline
is expected since galaxies in the cluster will gradually consume the
gas in their disks, and the possible sources of replenishment, such 
as HVC's or gas rich satellites, will be stripped away. However, a slow
decline is not adequate to explain the the strong Balmer absorption line
spectra frequently seen in the cluster galaxies (Couch \& Sharples, 1987,
Barger et al., 1995). 
In order to match the strength of such lines, a sudden decrease in the 
star formation rate is required (Poggianti \& Barbaro, 1996). 
In the more extreme cases, the line
strength can only be matched if the truncation is preceded by a burst 
of star formation; a burst would make the age distribution of the 
weaker lined systems easier to understand as well.

Galaxy harassment could provide the mechanism to initiate a burst of
star formation once a galaxy enters the cluster environment.  It is also
very efficient at causing instabilities that drive large amounts of gas
to the central regions of spirals (Lake \etal 1998), although
these process are less efficient in luminous spirals (Moore \etal 1999).
The numerical experiments of this paper put us in the position to assess
the plausibility of ram-pressure stripping as the truncation
mechanism. Initially, this scenario seems promising. In the Coma cluster
environment, the wind due to the ICM causes a substantial reduction
in the size of gaseous disks.  Indeed, Bothun \& Dressler (1986) find
several star-bursting, HI deficient spirals in the core of the Coma cluster.
The timescale for this is
very rapid and is shorter than the time taken to cross the cluster
core.  However, beyond this superficial success, a number of problems
remain to be addressed:

\begin{itemize}
\item The largest deficit of this model is that in no case is the
gas disk completely removed. A substantial portion of the cold 
gas remains sufficiently bound to the stellar disk such 
that the external medium
prefers to flow around the system. In the most extreme case, of a 
galaxy passing through the core of the Coma cluster at $ 3000\kms$,
the disk is truncated at $\sim 1.5$ disk scale lengths. 
We can estimate the corresponding reduction in the star-formation rate
using the Schmidt star formation law (Schmidt 1959, Kennicutt 1989) to 
calculate the contribution to the overall star formation
rate at each radius. For the unfortunate galaxy mentioned above, 
the star formation rate will be reduced by a factor 2.
In lower density environments, the effect
is much lower: stripping the disk beyond 3 scale lengths reduces the
star formation rate by only 10\%.
Fujita \& Nagashima (1998) recently examined the colour evolution
of spiral galaxies that have suffered ram-pressure stripping with
similar conclusions.

\item 
Several authors 
find that the star-formation rate in cluster galaxies
is significantly reduced between the field and the cluster center
(Dressler \etal 1997, Balogh \etal 1998, Poggianti \etal 1998).
Even when the diffuse gaseous material is stripped from the disk, 
additional gas will remain in the form of dense molecular clouds. 
These cannot be removed by the ram-pressure force since they 
are so small and dense. In local Sa-Sc galaxies, the mass of
molecular gas can equal the atomic gas fraction (Young \& Scoville, 1991).

\item Comparison of the stripped gas fractions for galaxies in the
cores of the Coma and Virgo cluster clusters shows that ram-pressure
is only a significant force in the densest regions. In contrast, the
data of Balogh \etal (1998), and of Morris \etal (1998), suggest that
the influence of the environment extends out to as much as twice the
cluster virial radius.  Some of this effect probably comes from
galaxies that are embedded in groups and poor clusters that are part of
the large-scale structure around the cluster. Secondly, galactic
orbits in clusters that form in a hierarchical universe are fairly
radial. Ghigna \etal (1998) demonstrate that 20\% of cluster galaxies
orbit with apocenter to pericenter ratios larger than 10:1. Thus, 20\%
of galaxies that have orbited through the core of the Coma cluster may
be found at, or beyond, the virial radius.  Nevertheless, to fully
explain this effect we require a mechanism that is effective in
environments less dense than the core of the Coma cluster.

\item Finally, we note that our simulations provide no explanation linking
the stripping of gas with a burst of star formation, as is required
to explain the most extreme absorption line spectra. The lack of such
a link most likely results from physical processes that have been 
omitted from our simulations. For instance, in the edge-on case,
the effect of the wind is to substantially compress the leading
edge of the disk. It is quite plausible, that this compression could 
lead to an increase in the collision rate of molecular clouds,
leading to a substantial enhancement in the star formation rate.
Fujita (1998) discusses the proposed mechanisms for inducing 
star-bursts in cluster galaxies, concluding that galaxy-harassment
is the most viable candidate.
\end{itemize}

This discussion suggests that simple ram-pressure stripping does not
adequately explain the sharp decline of star formation seen in
Butcher-Oemler galaxies. One possibility is that our models need to
be generalised to explicitly include the effects of star formation and
galaxy harassment.  This will tend to make galaxies more susceptible
to the ram pressure of the ICM. First because the molecular clouds
that are disrupted by star formation will not be able to re-form if
the diffuse material has already been removed from the
disk. Secondly, tidal shocks via galaxy harassment may tend to make
the disk structure more diffuse (and therefore more susceptible to
stripping); this process could be particularly important if the effect
of the stripping were to promote a burst of star formation.

We note that the restricted Eulerian treatment of this problem by 
Balsara \etal (1994) found that cooling gas may accrete back into the 
galaxy.  We do not observe this phenomenon, but this is due 
to our resolution in low density regions which are 
better resolved using grid based techniques. We are addressing 
this problem using higher resolution simulations performed using
parallel SPH and Eulerian codes.

\section{Conclusions}

We analyze the ram-pressure stripping process of a spiral galaxy
passing through the ICM using hydro-dynamical simulations and 
we conclude that:

$\bullet$ Ram pressure stripping is an effective mechanism at depleting gas 
from cluster spirals. The radius to which gas is removed can be calculated by
equating the ram pressure force $\rho v^2$ to the restoring force provided
by the disk, as originally suggested by Gunn \& Gott (1972).

$\bullet$ Bulges provide an additional gravitational force that dominate
the holding force in the central few kpc. Even a Milky Way type spiral crossing
the core of the Coma cluster at $3000\kms$ will retain gas within the central 
region.

$\bullet$ The time-scale for gas to be removed is very short $\sim
10^7$ years, a fraction of a crossing time, whereas the timescale for
gravitational interactions (galaxy harassment) to affect morphology 
and induce star-formation is of order a cluster crossing time.

$\bullet$ Disks moving through clusters with orbital inclination edge on to the
direction of motion lose about 50\% less gas than a full face on encounter
with the ICM.

$\bullet$ Observations of the $H_I$ distribution in cluster spirals show
evidence for tidally truncated disks by an amount roughly in accordance with
analytic expectations. 

$\bullet$ 
Ram pressure stripping alone does not provide the physical
mechanism behind the origin of the Butcher-Oemler effect. 

\section{Acknowledgments}

MGA would like to acknowledge support from {\it Fundaci\'on
Antorchas}, Argentina and the British Council. BM is supported by 
a Royal Society University Research Fellowship.

\end{document}